\documentclass [a4paper,12pt]{article}
\usepackage{amssymb}
\usepackage{graphicx}
\usepackage{cite}
\usepackage{amsmath}
\usepackage{fancybox}
\usepackage{color}
\usepackage{geometry}
\usepackage{mathtools}
 
\newcommand{\rme}{{\mathrm e}} 
\newcommand{\rmc}{{\mathrm c}} 
\newcommand{\rmM}{{\mathrm M}} 
\newcommand{\rmT}{{\mathrm T}} 
\newcommand{\rmD}{{\mathrm D}} 
\newcommand{\rmH}{\text{Heavy}}

\usepackage{tikz}
\usetikzlibrary{positioning,decorations,calc}

\begin{document}
\setlength{\baselineskip}{18pt}
\begin{titlepage}

\begin{flushright}

\end{flushright}
\vspace{0.3cm}
\begin{center}
{\LARGE\bf Conjugate Boundary Conditions, Kaluza-Klein Fermions, and an Extended Seesaw Model}
\end{center}
\vspace{5mm}

\begin{center}
{\large Yugo Abe,
Yuki Adachi$^a$, 
and Yukihiro Fujimoto $^{b}$ 
}
\end{center}

\vspace{0.3cm}

\centerline{{\it \small
Department of Pure and Applied Physics, Kansai University,
Suita 564-8680, Japan.
}}

\centerline{{\it\small
$^{a}$
Department of Sciences, Matsue College of Technology,
Matsue 690-8518, Japan.
}} 

\centerline{{\it \small
$^b$
Department of Integrated Arts and Science,
National Institute of Technology, Okinawa College,
Nago, 905-2192, Japan.
}}

\vspace{1.0cm}
\centerline{\large\bf Abstract}
\vspace{0.5cm}
In this paper, we discuss the conjugate boundary condition (CBC),
which has recently been studied as a way of realizing a Majorana fermion
within a compactified five-dimensional theory ($\mathcal M^4\otimes S^1/Z_2$).
The Majorana fermion which plays a crucial role in the seesaw scenario arises as a zero mode (lowest mode) by imposing the CBC.
Although each nonzero Kaluza-Klein (KK) mode is also naively expected to be described by two Majorana fermions, 
we show by direct calculation that they can be combined into a Dirac spinor in the free theory or up to the level of the quadratic term in the Lagrangian.
This difference comes from the fact that 
an accidental U(1) symmetry exists in the KK mode sector, though such a symmetry does not appear in the zero mode sector.
We also point out that the compatibility between the CBC and the axial U(1) transformation plays a crucial role in the diagonalization of the KK mode.

We also investigate interactions compatible with both the CBC and the
chiral orbifold projection.
We find that a nontrivial bulk interaction between a CBC fermion and a
chiral-orbifold fermion is allowed.
As an application, we construct an extended seesaw scenario by utilizing both boundary conditions and such nontrivial bulk interactions.
The resulting seesaw scenario has a different property in contrast with the conventional extended seesaw scenario; namely, 
the above new non-trivial interactions cause lepton number violation (LNV), while the Majorana mass term in our model does not violate lepton number.

\end{titlepage}

\section{Introduction} 
The Standard Model (SM) has been remarkably successful in describing the behavior of particles, including electroweak precision measurements. 
Despite such success of the SM,
the origin of the tiny neutrino mass is still a mystery in particle physics. 
The neutrino mass 
is much smaller than the masses of the charged leptons, in contrast to the quark sector,
so that there is a huge mass hierarchy in the lepton sector.
This suggests that there is another mechanism to reproduce the neutrino mass hierarchy.
The seesaw scenario is one of the well-known mechanisms that reproduce such a hierarchy,
and it is widely studied in several contexts.

In the seesaw scenario, once a large Majorana mass term for the right-handed neutrino is generated,
the active neutrino appears, and its mass becomes much smaller than the mass of a sterile neutrino.
In the simplest seesaw scenario (type-I) 
\cite{Minkowski:1977sc,Yanagida:1979as,Mohapatra:1979ia},
which introduces a Majorana mass term for the right-handed neutrino,
a large Majorana mass, such as the grand unified theory (GUT) scale, is typically needed.
There are also other tree-level realizations of the seesaw mechanism:
the type-II seesaw with a scalar triplet
\cite{Cheng:1980qt,Schechter:1980gr}
and the type-III seesaw with fermionic triplets
\cite{Foot:1988aq}.
Extensions of the type-I seesaw are called extended seesaw scenarios, which include the double seesaw, inverse seesaw, and linear seesaw
\cite{Mohapatra:1986aw,Akhmedov:1995ip}.
In these scenarios, additional neutral singlet fermions are introduced and are also widely discussed
because the required mass scale can be lower than the GUT scale.

Since the seesaw mechanism is an attractive scenario, 
it is natural to consider its compatibility with higher-dimensional theories\footnote{
Ref.~\cite{Fujimoto:2016} proposed a challenging extra-dimensional scenario, in which suitable boundary conditions make only the first Kaluza--Klein mode exponentially light, thereby identifying the small neutrino mass with a KK mass rather than with the Higgs-induced mass.
}. 
However, at first glance, the conventional Majorana-type seesaw seems to be
nontrivial in a five-dimensional theory because of the well-known fact that the conventional Majorana
condition cannot be imposed on a single five-dimensional spinor\footnote{The symplectic Majorana condition is another way to reproduce a Majorana fermion;
see, e.g., Refs.~\cite{VanProeyen:1999ni}. 
However, we do not discuss it in this paper.}.
In four-dimensional space-time, a Majorana fermion is defined by the identification between particle and antiparticle,
namely, 
$\psi^\rmc= \psi$,
where $\psi^\rmc\coloneq C\bar{\psi}^{\rmT}=-i\gamma_2\psi^\ast$ in the chiral representation.
For five-dimensional space-time, the charge conjugation operation has a different
property from the four-dimensional case, 
\begin{equation}
  \psi^{5\rmc}\coloneq C_{5}\bar{\psi}^{\rmT}=-i\gamma_5\gamma_2\psi^\ast =\gamma_5\psi^\rmc,
\end{equation}
so that the ordinary Majorana condition
for a single spinor cannot be imposed in five-dimensional space-time because $(\psi^{5\rmc})^{5\rmc}=-\psi$.
The $\gamma_5$ in $\psi^{5\rmc}$ guarantees the invariance of the kinetic term along the extra dimension.
In other words, the charge conjugation in five dimensions includes four-dimensional charge conjugation and a chiral transformation.
However, this property conflicts with the Majorana condition, so that the conventional Majorana fermion does not exist in five-dimensional space-time.

In order to realize the seesaw mechanism in higher-dimensional space-time,
there are two possible strategies.
One is to introduce Majorana fermions as brane-localized fermions.
This is possible because the fixed point or the brane is treated as a
four-dimensional space-time.
Another strategy is to impose a conjugate boundary condition (CBC)
\cite{Grzadkowski:2005yz,Abe:2016vfs}
on the bulk fermions
\footnote{Similar charge-conjugating boundary conditions have also been discussed
under the name of pseudo-Majorana boundary conditions in warped
extra-dimensional models~\cite{Brakke:2008zb}.}
.
Roughly speaking, the CBC is a boundary condition that identifies a particle
with its antiparticle under the reflection of the extra-dimensional coordinate,
$y\to -y$.
Then, the zero mode can be identified as a four-dimensional Majorana fermion.
This mechanism is analogous to the usual chiral orbifolding.
Although a chiral fermion does not exist as a bulk fermion in five-dimensional
space-time, a chiral zero mode can arise after imposing the orbifold projection.
Then, the CBC opens a new avenue to construct the seesaw scenario
using only bulk fields.
A typical CBC on a five-dimensional fermion in the $\mathcal M^4\otimes S^1/Z_2$ space-time with radius $R$ is written as
\begin{equation}
\label{eq:CBC-intro}
  \psi(y_i-y)=P\psi^\rmc(y_i+y),   ~~~~~(i=0,1)
\end{equation}
where $y_{0,1}=0,\pi R$ and $P$ stands for the parity of the fields ($P^2=1$).
Namely, the Majorana condition is realized through this boundary condition rather than as a bulk condition on the five-dimensional spinor.
We emphasize that $\psi^\rmc$ represents the four-dimensional charge conjugation of $\psi$.
We note that the charge conjugation in the CBC is the four-dimensional one in order to keep the kinetic term consistent with the reflection $y\to-y$, as discussed in Sec.~\ref{sec:CBC}. 
In this paper,
we solve the bulk equations and explicitly derive Kaluza-Klein (KK) mass eigenstates which 
are consistent with Ref.~\cite{Grzadkowski:2005yz},
as presented in Sec.~\ref{sec:modefunctionCBC}.
In our result, in the free theory, 
each nonzero KK level consists of two degenerate Majorana fermions
so that they are described in terms of a Dirac spinor rather than a Majorana fermion.
In fact, we show that there is an accidental U(1) symmetry in each nonzero KK mode.
Such degeneracy may be lifted by interactions, for example, by the vacuum expectation value (VEV) of $A_y$ in the context of the gauge-Higgs unification scenario,
as pointed out in Ref.~\cite{Grzadkowski:2005yz}.
In other words, this indicates that this accidental U(1) symmetry is broken by the VEV of $A_y$.

The CBC \eqref{eq:CBC-intro} is not invariant under the vector-like U(1) transformation
\(\psi\to e^{i\alpha}\psi\). 
Therefore, a definite fermion number cannot be assigned to the CBC fermion.
On the other hand, the CBC is invariant under the axial U(1)
transformation \(\psi\to e^{i\alpha\gamma_5}\psi\).
This is consistent with the fact that the axial U(1) transformation preserves the Majorana condition.
This property plays an important role in the diagonalization of the KK
mass eigenstates, as discussed in Sec.~\ref{sec:modefunctionCBC}.
We also find that non-trivial bulk interactions between fermions with the CBC and chiral orbifolding are allowed:
\begin{equation}
\label{eq:nontrivialbulkinteraction}
  M_N\bar\eta\left[\psi - PP' \gamma_5 \psi^\rmc\right]
  +\text{(h.c.)}
\end{equation}
where $\psi$ and $\eta$ stand for fermions with the CBC and chiral orbifolding, respectively.
The field $\eta$ satisfies
\begin{equation}
  \eta(y_i-y)=P'\gamma_5\eta(y_i+y)
\end{equation}
where $P'$ stands for the parity.
The non-trivial interaction \eqref{eq:nontrivialbulkinteraction} is consistent with five-dimensional Lorentz invariance because the second term in the square brackets can be written in terms of five-dimensional charge conjugation 
$\gamma_5 \psi^\rmc = \psi^{5\rmc}$.
As a result, this non-trivial bulk interaction allows us to construct an extended seesaw
scenario using only bulk fermions.
After compactification, the zero-KK-mode sector realizes an extended seesaw
mass matrix, as discussed in Sec.~\ref{sec:SEESAW}.
We also note that the CBC fermion $\psi$ does not carry lepton
number because the CBC is not invariant under the vector-like U(1) transformation.
Therefore, unlike in conventional extended seesaw scenarios, lepton number violation (LNV) in our model is induced by the above bulk interaction rather
than being introduced directly as a Majorana-type mass term in the neutral-fermion mass matrix.

The main results of this paper are threefold.
First, we explicitly derive the mode functions and KK mass eigenstates of a CBC fermion with a constant bulk mass.
Second, we show that the zero mode is a four-dimensional Majorana fermion, whereas each nonzero KK mode forms a Dirac fermion in the free theory.
Third, we find non-trivial bulk interactions compatible with both the CBC and the chiral orbifold projection. As an application, we construct an all-bulk extended seesaw model. 

This paper is organized as follows.
In Sec.~\ref{sec:CBC}, we introduce the conjugate boundary condition and construct a possible interaction between a CBC fermion and a chiral-orbifold fermion.
In Sec.~\ref{sec:modefunctionCBC}, we derive the KK mass eigenstates and mode functions of a CBC fermion precisely.
In Sec.~\ref{sec:SEESAW}, we provide a new model of the extended seesaw scenario with the CBC and chiral orbifolding.
The phenomenological implications of this scenario are also discussed.
Finally, we give a summary and discussion.

\section{Conjugate boundary condition and allowed interactions} 
\label{sec:CBC}
\subsection{Conjugate boundary condition}
In this section, we review the conjugate boundary condition that gives rise to a four-dimensional Majorana fermion in five-dimensional space-time ($\mathcal M^4\otimes S^1/Z_2$)
\cite{Grzadkowski:2005yz,Abe:2016vfs}.
The CBC assigns opposite parities to the real and imaginary parts of a five-dimensional spinor field \(\psi(y)\).
The explicit form of the CBC was shown in the Introduction; we show it again:
\begin{equation}
  \label{eq:4DCBC}
  \psi(y_i -y) = P \psi^\rmc (y_i+y) 
\end{equation}
where $\psi^\rmc$ stands for the four-dimensional charge conjugation of the spinor field.
Here, $R$ is the radius of $S^1$ and $y_i$ are the fixed points $y_0=0,y_1=\pi R$.
The parameter $P=\pm 1$ is the parity of the singlet fermion; it becomes a parity matrix for a multiplet fermion
\footnote{This boundary condition corresponds to $P1-R1$ in Ref.~\cite{Grzadkowski:2005yz};
we point out that an additional parity can be assigned in this situation.}
.
In general, the parities at the fixed points can be chosen differently. However, we assume that the parities are the same at both fixed points for simplicity, which can be achieved by imposing periodicity on $\psi$.
The field $\psi^\rmc$ is the charge conjugate of $\psi$, so that the zero-mode sector is expected to become a four-dimensional Majorana fermion.

As mentioned in the Introduction, we emphasize that the charge conjugation operation in Eq.~\eqref{eq:4DCBC} is the four-dimensional one.
It is worth explaining why the four-dimensional charge conjugation is necessary for this boundary condition.
At first glance, one might expect the suitable CBC to adopt the five-dimensional charge conjugation as
\begin{equation}
  \label{eq:5DCBC}
  \psi(y_i -y) = P \psi^{5\rmc} (y_i+y) 
\end{equation}
from a naive guess. 
This is because the five-dimensional charge conjugation preserves the kinetic term 
\begin{equation}
 \bar \psi i\partial_M\Gamma^M \psi \overset{C_{5}}{\to} \bar \psi i\partial_M\Gamma^M \psi,\label{180250_28Jun26}
\end{equation}
where the index $M$ runs over $\mu=0,1,2,3$ and $y$.
The five-dimensional gamma matrices $\Gamma^M$ obey the five-dimensional Clifford algebra $\{\Gamma^M,\Gamma^N\}=2\eta^{MN}$ ($\eta^{MN}=\text{diag}(+,-,-,-,-)$);
$\Gamma^\mu=\gamma^\mu$ and $\Gamma^y=i\gamma^5=i\gamma_5=\gamma^{0}\gamma^{1}\gamma^{2}\gamma^{3}$.

However, it does not preserve the kinetic term along the $y$-direction.
The $y$-derivative $\partial_y$ flips sign under the reflection $y\to-y$ as $\partial_y\to -\partial_y$, so the kinetic term changes
as
\begin{equation}
\bar\psi i\partial_y\Gamma^y\psi \to -\bar\psi i\partial_y\Gamma^y\psi,
\end{equation}
under Eq.~\eqref{eq:5DCBC}.
In order to preserve the kinetic term $\bar \psi i\partial_y \Gamma^y \psi$, a chiral rotation should be performed on the right-hand side of Eq.~\eqref{eq:5DCBC}
\footnote{The necessity of $\gamma_5$ is the same as in ordinary chiral orbifolding, as shown in Eq.~\eqref{eq:BC-CBCandCOP}.}
,
so the modified boundary condition is 
\begin{equation}
\label{eq:5DCBCmodefied}
  \psi(y_i -y) =\gamma_5 P \psi^{5\rmc} (y_i+y).
\end{equation}
The $\gamma_5\psi^{5\rmc}$ in the above expression is nothing but the four-dimensional charge conjugation, so that it is completely equivalent to the CBC \eqref{eq:4DCBC}.
In fact, the five-dimensional charge conjugation is $\psi^{5\rmc} =-i\gamma_5\gamma_2\psi^\ast$ in the chiral representation; therefore, the right-hand side of Eq.~\eqref{eq:5DCBCmodefied} becomes $\gamma_{5}\psi^{5\rmc}=\psi^{\rmc}$.

Note that 
the CBC clearly conflicts with the vector-like U(1) transformation, although the axial U(1) transformation is compatible with the CBC.
This suggests that such a global U(1) symmetry is broken by the boundary condition.
This is consistent with the fact that the zero mode of $\psi$ becomes a Majorana fermion,
\emph{i.e.},
a Majorana fermion cannot be assigned any charge.
For the axial U(1) transformation, we point out that it is compatible with the CBC due to the fact that both $\psi$ and $\psi^\rmc$ transform in the same way under the chiral U(1) transformation:
\begin{equation}
\label{eq:BC-CBCandCOP}
  \psi\to \rme^{i\alpha\gamma_5} \psi,
  \psi^\rmc\to \rme^{i\alpha\gamma_5} \psi^\rmc,
\end{equation}
or equivalently, 
\begin{equation}
  \psi_\rmM\to \rme^{i\alpha\gamma_5} \psi_\rmM
\end{equation}
where $\psi_\rmM \coloneq\frac{1}{\sqrt2}[\psi+\psi^\rmc]$.
This suggests that the Majorana condition $\psi=\psi^\rmc$ is preserved after the chiral U(1) transformation.
This property plays a crucial role in the diagonalization of the nonzero KK modes, as we will discuss in the following sections, 
because the kinetic term in the $y$-direction behaves as a pseudo-scalar mass, so that the 
axial U(1) transformation is necessary for the diagonalization.

\subsection{Allowed interactions with the CBC and chiral orbifold projection}
In this subsection, we discuss typical interactions that are allowed in the framework of five-dimensional space-time ($\mathcal M^4\otimes S^1/Z_2$, whose radius is given by $R$).
The CBC and chiral orbifold projection, together with periodicity conditions on the fields, are given by 
\begin{equation}
\label{eq:boundarycondition}
  \psi (y_i-y) = P\psi^\rmc (y_i+y)
  ,
  \eta (y_i-y) = P'\gamma_5\eta (y_i+y),
\end{equation}
where $y_0=0,y_1=\pi R$.
Note that the same parities at the two fixed points are imposed by the periodicity.
First, the mass term for $\eta$ is allowed by five-dimensional Lorentz invariance, but it is not compatible with chiral orbifolding. 
That is why the odd-sign bulk mass term $\epsilon(y)\bar\eta\eta$ is introduced, as is well known in chiral orbifolding theory\footnote{For a neutral fermion, the bilinear $\bar\eta\eta^{5\rmc}+\text{(h.c.)}$ is compatible with the chiral
orbifold projection. 
Such bulk Majorana mass terms for gauge-singlet five-dimensional neutrinos have been discussed, for example, in \cite{Watanabe:2010mnta}. 
In the present paper, however, we do not introduce this term for the chiral-orbifold fermion $\eta$. 
Instead, the Majorana mass appearing in the low-energy theory is generated from the ordinary CBC-compatible bulk mass term for the CBC fermion, 
as discussed below.
}.

As for the CBC, in contrast to the chiral orbifold projection,
the mass term $m\bar\psi\psi$ is compatible with five-dimensional Lorentz invariance and the CBC
since
\begin{equation}
 m\bar\psi(y)\psi(y) 
  \to 
  m\overline{\psi}(-y)\psi(-y)=m\overline{\psi^{\rmc}}(y)\psi^{\rmc}(y)=m\bar\psi(y)\psi(y).
\end{equation}
Thus, the Lagrangian up to quadratic order with the CBC is written as
\begin{equation}
  \mathcal L = \bar\psi(i\partial_M\Gamma^M -m)\psi.
\end{equation}
We will derive the mode functions and their KK mass eigenstates in the next subsection.

In this paper, we discuss the two different kinds of boundary conditions
\eqref{eq:boundarycondition}.
In the five-dimensional sense,
the quadratic interactions $\bar\eta \psi$ and $\bar\eta \gamma_5\psi^\rmc$ are allowed; however, they are not compatible with the boundary conditions.
Therefore, the following 
non-trivial
quadratic interactions with neutral fermions can be introduced
in the context of the five-dimensional theory:
\begin{equation}
  A\bar \eta (\psi \pm\psi^{5\rmc}) + \text{(h.c.)}
  =
  A\bar \eta (\psi \pm \gamma_5\psi^{\rmc}) + \text{(h.c.)}.
\end{equation}
The five-dimensional charge conjugation is written in terms of the four-dimensional one; 
$\psi^{5\rmc}=\gamma_5 \psi^{\rmc}$, as mentioned in the Introduction.
We note that no charge can be assigned to $\psi$ or $\eta$, because the above interaction does not preserve the charge.
In order to derive the above interaction, we begin with the following linear combination:
\begin{equation}
  A\bar \eta \psi -B\bar \eta \gamma_5\psi^{\rmc}.
\end{equation}
The coefficients $A$ and $B$ cannot be independent of each other due to the 
boundary conditions 
\eqref{eq:boundarycondition}.
In fact, after the reflection $y\to -y$, it becomes
\begin{equation}
  A\bar \eta \psi -B\bar \eta \gamma_5\psi^{\rmc}
  \to
  A\bar \eta' \psi' -B\bar \eta' \gamma_5\psi'{}^{\rmc}
  =
  P'PB\bar \eta \psi-P'PA\bar \eta \gamma_5\psi^\rmc .
\end{equation}
This indicates that they satisfy
$A=PP'B$,
so the allowed interaction should be
\begin{equation}
\label{eq:majoranainteraction}
  A\bar\eta(\psi-P'P\gamma_5\psi^\rmc).
\end{equation}
Such a non-trivial interaction enables us to construct an extended seesaw scenario in Sec.~\ref{sec:SEESAW}.

The above interaction can be extended to an interaction with a charged scalar field $\phi$, such as $\phi\bar\eta \psi$.
Such an interaction can be written as
\begin{equation}
A \bar\eta(\phi^\ast\psi-PP'P''\phi\gamma_5\psi^\rmc)
\end{equation}
by the same procedure.
In this case, both $\psi$ and $\phi$ can carry some charge, while $\eta$ remains neutral.
Note that the scalar field $\phi$ should have exactly the same charge as $\psi$ and obey a similar CBC:
\begin{equation}
  \phi(y_i-y) = P''\phi^\ast (y_i+y).
\end{equation}
After dimensional reduction, the zero mode of the scalar field $\phi$ becomes neutral in the same sense as $\psi$
\footnote{This mechanism was already pointed out in Ref.~\cite{Haba:2008ar}.}
.
Namely, the above CBC extracts the real or imaginary part of $\phi$ by assigning the parity $P''=+1$ or $P''=-1$, respectively.
We note that $\psi$ and $\phi$ have the same charge, but the remaining zero mode becomes neutral due to the CBC.
Such an extension is a candidate for the ultraviolet (UV) completion of
\eqref{eq:majoranainteraction}, as we will discuss in Sec.~\ref{sec:SEESAW}.

\section{Mode functions and KK mass eigenstates in the CBC}
\label{sec:modefunctionCBC}
In this section, we discuss the mode functions of a fermion satisfying the CBC. 
We begin with the following action:
\begin{equation}
  S
  =\int d^4x \int_{-\pi R}^{\pi R}dy ~ \mathcal L
  =\int d^4x \int_{-\pi R}^{\pi R}dy ~ \bar \psi (i\partial_M\Gamma^M -m)\psi.\label{205930_28Jun26}
\end{equation}
In the CBC, the constant bulk mass term is allowed in contrast to chiral orbifolding,
as we discussed in the previous section.
Since there is no discontinuous term in the Lagrangian, 
the mode functions are expected to be trigonometric functions.
Now we define the Majorana fermions as 
\begin{equation}
  \psi_\rmM \coloneq\frac{1}{\sqrt2} (\psi+\psi^\rmc),
  \psi_{\bar\rmM} \coloneq\frac{1}{\sqrt2 i} (\psi-\psi^\rmc),\label{204611_28Jun26}
\end{equation}
which correspond to the real and imaginary parts of the spinor field.
Then, by choosing the parity as $P=1$, the mode expansion with the following CBC
\begin{equation}
\label{eq:CBC}
  \psi(y_i -y) = + \psi^\rmc (y_i+y) 
\end{equation}
is described by
\begin{equation}
  \psi(y)=\frac{1}{\sqrt2}
  \sum_{n=0}^\infty\left[\psi_{\rmM}^{(n)}C_n(y)+i\psi_{\bar\rmM}^{(n)}S_n(y) \right]
\end{equation}
or equivalently,
\begin{equation}
  \psi_\rmM(y) = \frac{\psi(y)+\psi^{\rmc}(y)}{\sqrt2} =\sum_{n=0}^\infty \psi_{\rmM}^{(n)} C_n(y)\label{204515_28Jun26}
\end{equation}
and
\begin{equation}
  \psi_{\bar\rmM}(y) = \frac{\psi(y)-\psi^{\rmc}(y)}{\sqrt2 i} =\sum_{n=0}^\infty \psi_{\bar\rmM}^{(n)} S_n(y)\label{204315_28Jun26}
\end{equation}
where $C_n(y)$ and $S_n(y)$ are even and odd functions of $y$, respectively.
We again note that periodicity is imposed on $\psi$, so that the same parity is assigned at the two fixed points, as shown in \eqref{eq:CBC}.
The equation of motion derived from the action is given by
\begin{equation}
  (i\partial_M\Gamma^M -m)\psi(y)=0.
\end{equation}
As we will see later, the $y$-derivative mixes $\psi_{\rmM}$ and $\psi_{\bar\rmM}$ with each other.
Thus, we move to the second-order differential equation, namely, the Klein-Gordon equation.
Multiplying by $i\partial_M\Gamma^M +m$,
the Klein-Gordon equation is obtained:
\begin{equation}
  (\partial_M \partial^M+m^2)\psi(y)=0.
\end{equation}
Since the Klein-Gordon equation is second order in $y$, it does not mix
the even and odd components. Therefore, after imposing the CBC, the even
component $\psi_{\rmM}$ and the odd component $\psi_{\bar{\rm M}}$ can be
expanded by $C_n(y)$ and $S_n(y)$, respectively.
After substituting the mode expansion and replacing $i\partial_\mu\gamma^\mu$ with the mass eigenvalues $m_n$ using the four-dimensional Dirac equation
$(i\partial_\mu\gamma^\mu -m_n)\psi_{\rmM}^{(n)}=0$,
we have 
\begin{equation}
  \sum_{n=0}^\infty (m_n^2 +\partial_y^2 -m^2)\psi^{(n)}_{\rmM}C_n(y)=0.
\end{equation}
Since the mode function diagonalizes the KK modes, it satisfies
\begin{equation}
  [m_n^2-m^2+\partial_y^2]C_n(y)=0.
\end{equation}
As the mode function $C_n(y)$ is even, the solution is proportional to
$\cos k_ny$.
By imposing the CBC
 $\psi_\rmM(y_i-y)=\psi_\rmM(y_i+y)$,
 we have
 \[\cos k_n(\pi R-y)=\cos k_n(\pi R+y).\]
 This implies $k_n=\frac{n}{R}$, so that $C_n(y)$ becomes 
\begin{equation}
  C_n(y) \propto \cos \frac{ny}{R}
\end{equation}
and its KK mass eigenvalue is given by
\begin{equation}
  m_n=\sqrt{m^2+\left(\frac{n}{R}\right)^2}.
\end{equation}
This is consistent with the result that we will obtain later.
The same procedure can be applied to $\psi_{\bar \rmM}$, which has odd parity, 
and we have 
\begin{equation}
  S_n(y)\propto \sin \frac{ny}{R}
\end{equation}
with the same KK mass eigenvalue $m_n$.
Then the normalized mode functions are given as follows:
\begin{equation}
    C_n(y) = \frac{1}{\sqrt{2^{\delta_{n0}}\pi R}} \cos\left(\frac{ny}{R}\right), \quad S_n(y) = \frac{1}{\sqrt{\pi R}} \sin\left(\frac{ny}{R}\right).
\end{equation}
They form an orthogonal set on the domain $[-\pi R,\pi R]$ as 
\begin{equation}
    \int_{-\pi R}^{\pi R} dy \, C_n C_m =  \int_{-\pi R}^{\pi R} dy \, S_n S_m = \delta_{nm}, \quad \int_{-\pi R}^{\pi R} dy \, C_n S_m = 0.
\end{equation}
Finally, we have the mode expansion of $\psi(y)$:
\begin{equation}
  \psi(y)
  =\frac{1}{\sqrt{2}}
  \sum_{n=0}^\infty\left[\psi_{\rmM}^{(n)}C_n(y)+i\psi_{\bar\rmM}^{(n)}S_n(y) \right]
  =\frac{1}{\sqrt{2}}
  \psi_{\rmM}^{(0)}C_0(y)
  +\frac{1}{\sqrt{2}}
  \sum_{n=1}^\infty\left[\psi_{\rmM}^{(n)}C_n(y)+i\psi_{\bar\rmM}^{(n)}S_n(y) \right].
\end{equation}
We note that each nonzero KK mode is degenerate; namely, $\psi^{(n)}_\rmM$ and $\psi^{(n)}_{\bar\rmM}$ have the same KK mass eigenvalue $m_n$,
so that they can be combined into a Dirac spinor, as shown later.

Next we discuss the nonzero KK mass eigenstates. 
After substituting the above mode functions and by utilizing 
\begin{equation}
  \partial_y C_n(y)=-\frac{n}{R}S_n(y),
  \partial_y S_n(y)=\frac{n}{R}C_n(y),
\end{equation}
we have
\begin{equation}
\label{eq:reducedquadraticterm}
  \begin{split}
    &\int_{-\pi R}^{\pi R} dy~ \bar\psi (y)[i\partial_\mu \gamma^\mu -\partial_y\gamma_5-m]\psi(y)
    \\
    &=\frac12 \overline{\psi_{\rmM}^{(0)}}(i\partial_\mu\gamma^\mu -m)\psi^{(0)}_{\rmM}
    +\sum_{n=1}^\infty\frac{1}{2} 
      \overline{\Psi^{(n)}_\rmM}
   \left[
   i\partial_{\mu}\gamma^{\mu}\cdot \boldsymbol{1} -
   \begin{pmatrix}
    m& \frac{n}{R}i\gamma_{5}\\
    \frac{n}{R}i\gamma_{5}& m
   \end{pmatrix}
   \right]
   \Psi^{(n)}_\rmM
  \end{split}
\end{equation}
where we employ the abbreviation $\Psi_\rmM ^{(n)}=\begin{pmatrix}\psi_\rmM^{(n)},\psi_{\bar\rmM}^{(n)}\end{pmatrix}^\rmT$.
The zero mode sector is completely diagonalized, 
our task is diagonalization of the nonzero KK sector so that we concentrate on them.
Before precise derivation, we put some comments below.
The mass matrix term in \eqref{eq:reducedquadraticterm} contains the mixing between \(\psi_{\mathrm{M}}\) and \(\psi_{\bar{\mathrm{M}}}\), which is essentially the mixing between \(\psi\) and \(\psi^{\rmc}\), as a consequence of the CBC which transforms \(\psi\) to \(\psi^{\rmc}\) in the reflection \(y\to -y\).
Namely, the identification between $\psi$ and $\psi^\rmc$ while in the reflection $y\to -y$ leads such mixings\footnote{It is known as a Nambu basis.}.
We also note that the nonzero KK mode is ordinary four-dimensional Dirac spinor rather than an eight-component spinor. 
It is because the $\psi$ and $\psi^\rmc$ are not independent of each other.
In other words, the ingredients of $\Psi_\rmM^{(n)}$ are two independent Majorana spinors.
It indicates that if the mass eigenvalues are degenerate after diagonalization, then they are described by the Dirac spinor.

The mass matrix is symmetric, so it can be diagonalized by a suitable unitary matrix including the axial U(1) transformation
by utilizing the following matrix 
with the angle $\tan2\theta_n=\frac{n/R}{m}$:
\begin{equation}
    \begin{bmatrix} 
     m & \frac{n}{R}i\gamma_5 \\ 
    \frac{n}{R}i\gamma_5&  m 
    \end{bmatrix} 
    =m_n\rme^{2i\theta_n \bar\Gamma_5}
    ,
  \bar\Gamma_5 =
    \begin{bmatrix} 
     0 & \gamma_5 \\ 
    \gamma_5 &  0 
    \end{bmatrix} 
\end{equation}
where $m_n=\sqrt{m^2+\left(\frac{n}{R}\right)^2}$.
Then we have
\begin{equation}
    \sum_{n=1}^\infty\frac{1}{2} 
      \overline{\Psi^{(n)}_\rmM}
    \begin{bmatrix} 
    i\partial_{\mu}\gamma^{\mu} - m & -\frac{n}{R}i\gamma_5 \\ 
    -\frac{n}{R}i\gamma_5 & i\partial_{\mu}\gamma^{\mu} - m 
    \end{bmatrix} 
      \Psi^{(n)}_\rmM
    =
  \sum_{n=1}^\infty \frac{1}{2} 
      \overline{\Psi'^{(n)}_\rmM}
      (i\partial_\mu\gamma^\mu-m_n)
      \Psi'^{(n)}_\rmM
\end{equation}
where $\Psi'^{(n)}_\rmM = \rme^{i\theta_n \bar\Gamma_5}\Psi_\rmM^{(n)}$.
As mentioned in the Introduction, the Majorana condition is preserved after the axial U(1) transformation,
so that $\Psi'^{(n)}_\rmM$ is still a Majorana fermion.
Since the KK masses are degenerate at each level, they are described by a single Dirac spinor.
In fact, the expression obtained above can be rewritten in the form:
\begin{equation}
\label{eq:diagolalizedKK}
  \sum_{n=1}^\infty \frac{1}{2} 
      \begin{pmatrix}
        \overline{\psi'^{(n)}_\rmM},\overline{\psi'^{(n)}_{\bar \rmM}}
      \end{pmatrix}
      (i\partial_\mu\gamma^\mu -m_n)
      \begin{pmatrix}
        \psi'^{(n)}_\rmM\\ \psi'^{(n)}_{\bar \rmM}
      \end{pmatrix}.
\end{equation}
This form clearly shows that each nonzero KK mode consists of two independent Majorana fermions with the same mass eigenvalue $m_n$. 
Then, such two degenerate Majorana fermions are equivalently described by a single Dirac spinor
$\Psi^{(n)}_\rmD=\frac{1}{\sqrt2}(\psi'^{(n)}_\rmM + i\psi'^{(n)}_{\bar \rmM})$, 
namely,
\begin{equation}
\frac12
      \begin{pmatrix}
        \overline{\psi'^{(n)}_\rmM},-i\overline{\psi'^{(n)}_{\bar \rmM}}
      \end{pmatrix}
      (i\partial_\mu\gamma^\mu -m_n)
      \begin{pmatrix}
        \psi'^{(n)}_\rmM\\ i\psi'^{(n)}_{\bar \rmM}
      \end{pmatrix}
      =
      \overline{\Psi^{(n)}_\rmD}(i\partial_\mu\gamma^\mu -m_n)\Psi^{(n)}_\rmD.
\end{equation}
Finally, we obtain
\begin{equation}
  \begin{split}
  \int_{-\pi R}^{\pi R}dy ~ \mathcal L =&
   \frac12 \bar\psi_{\rmM}^{(0)}(i\partial_\mu\gamma^\mu -m)\psi^{(0)}_{\rmM}
   +\sum_{n=1}^\infty
   \overline{\Psi^{(n)}_\rmD} (i\partial_\mu \gamma^\mu-m_n)
   \Psi^{(n)}_\rmD.
  \end{split}\label{205903_28Jun26}
\end{equation}
Namely, the nonzero KK modes form Dirac spinors, though the zero mode is still written in terms of a Majorana fermion.
We note that this result is valid up to quadratic order in the action or for a completely free system.
If the action contains interactions with other fields, such as gauge interactions, the nonzero KK modes may be distinguished by the interaction.
In fact, if the gauge interaction is switched on, the KK modes split into two Majorana fermions 
because of the vacuum expectation value (VEV) of $A_y$, or equivalently the Scherk-Schwarz mechanism,
as shown in Ref.~\cite{Grzadkowski:2005yz}.

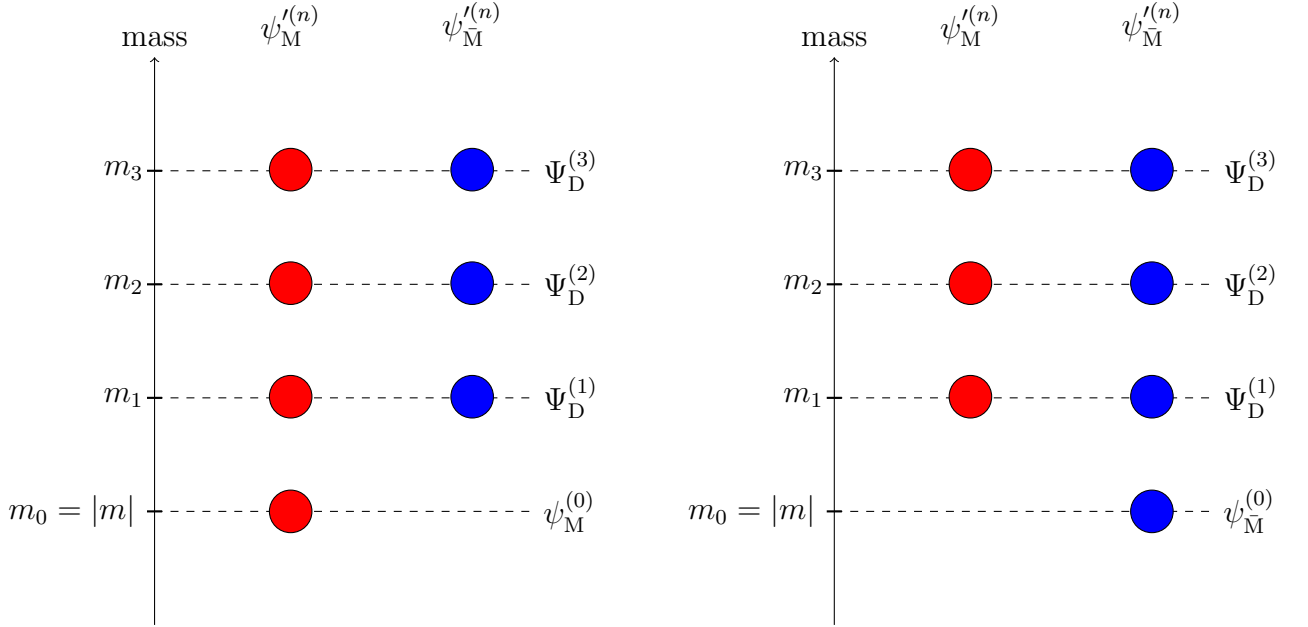
\begin{figure}[h]
\centering
\begin{tikzpicture}
\coordinate(O) at (0,0);
\coordinate(ey) at (0,1.5);
\coordinate(ex) at (1,0);
\coordinate(dy) at ($0.01*(ey)$);
\coordinate(dx) at ($0.1*(ex)$);
\coordinate(Dx) at ($4*(ex)$);

\coordinate(M) at ($0.2*(Dx)+(ex)$);
\coordinate(barM) at ($0.8*(Dx)+(ex)$);

\coordinate(Y) at ($5*(ey)$);

\path []($(M)+5*(ey)$) node[above]{$\psi'^{(n)}_\rmM$};
\path []($(barM)+5*(ey)$) node[above]{$\psi'^{(n)}_{\bar\rmM}$};

\path[draw, ->] (O) --(Y) node [above]{mass};

\path [draw, thick]($(ey)-(dx)$)node[left]{$m_0=|m|$}--+($2*(dx)$);
\path [draw, dashed,thin]($(ey)$)--+($(Dx)+(ex)$)node[right]{$\psi^{(0)}_\rmM$};

\path [draw ,fill=red ]($(ey)+(M)$) circle(8pt);

\foreach \x in { 1, 2, 3 } {
\path [draw, thick]($\x*(ey)+(ey)-(dx)$)--+($2*(dx)$);
\path [draw, dashed,thin]($\x*(ey)+(ey)$)node[left]{$m_\x$}--+($(Dx)+(ex)$)node[right]{$\Psi_\rmD^{(\x)}$};

\path [draw ,fill=red ]($\x*(ey)+(ey)+(dy)+(M)$) circle(8pt);
\path [draw ,fill=blue ]($\x*(ey)+(ey)+(dy)+(barM)$) circle(8pt);
}

\end{tikzpicture}~~~~~~~~~
\begin{tikzpicture}
\coordinate(O) at (0,0);
\coordinate(ey) at (0,1.5);
\coordinate(ex) at (1,0);
\coordinate(dy) at ($0.01*(ey)$);
\coordinate(dx) at ($0.1*(ex)$);
\coordinate(Dx) at ($4*(ex)$);

\coordinate(M) at ($0.2*(Dx)+(ex)$);
\coordinate(barM) at ($0.8*(Dx)+(ex)$);

\coordinate(Y) at ($5*(ey)$);

\path []($(M)+5*(ey)$) node[above]{$\psi'^{(n)}_\rmM$};
\path []($(barM)+5*(ey)$) node[above]{$\psi'^{(n)}_{\bar\rmM}$};

\path[draw, ->] (O) --(Y) node [above]{mass};

\path [draw, thick]($(ey)-(dx)$)node[left]{$m_0=|m|$}--+($2*(dx)$);
\path [draw, dashed,thin]($(ey)$)--+($(Dx)+(ex)$)node[right]{$\psi^{(0)}_{\bar\rmM}$};

\path [draw ,fill=blue]($(ey)+(barM)$) circle(8pt);

\foreach \x in { 1, 2, 3 } {
\path [draw, thick]($\x*(ey)+(ey)-(dx)$)--+($2*(dx)$);
\path [draw, dashed,thin]($\x*(ey)+(ey)$)node[left]{$m_\x$}--+($(Dx)+(ex)$)node[right]{$\Psi_\rmD^{(\x)}$};

\path [draw ,fill=red ]($\x*(ey)+(ey)+(dy)+(M)$) circle(8pt);
\path [draw ,fill=blue ]($\x*(ey)+(ey)+(dy)+(barM)$) circle(8pt);
}

\end{tikzpicture}
\caption{
Schematic picture of the four-dimensional KK mass spectrum under the conjugate boundary condition. 
The mass eigenvalue $m_n$ is given by $m_n=\sqrt{m^2+\left(\frac{n}{R}\right)^2}$.
The left panel shows the four-dimensional mass spectrum of a free five-dimensional fermion under the CBC with parity $P=1$.
Due to the CBC, the zero mode appears as a four-dimensional Majorana fermion $\psi_\rmM^{(0)}$, whereas at each nonzero KK mode the two Majorana fermions have degenerate masses and form a Dirac fermion. 
The right panel shows the four-dimensional mass spectrum for the parity $P=-1$,
where the remaining zero mode is $\psi_{\bar\rmM}^{(0)}$.
}
\label{fig:CBC-spectrum}
\end{figure}

In our case, 
the choice of the parity $P$ determines the remaining zero-mode fermion, $\psi_\rmM^{(0)}$ or $\psi_{\bar\rmM}^{(0)}$. 
An illustrative figure is shown in Fig.~\ref{fig:CBC-spectrum}.
The left panel in Fig.~\ref{fig:CBC-spectrum} shows the four-dimensional mass spectrum of a free five-dimensional fermion under the CBC with parity $P=1$.
Due to the CBC, the zero mode appears as a four-dimensional Majorana fermion, whereas at each nonzero KK mode the two Majorana fermions have degenerate masses and form a Dirac fermion. 
The right panel shows the four-dimensional mass spectrum for the parity $P=-1$.
In such a case, the zero-mode fermion becomes $\psi_{\bar \rmM}^{(0)}$.
We also note that the mass of the zero mode becomes a Majorana mass term, although it originates from the ordinary five-dimensional mass term in the bulk Lagrangian.

Let us comment on the relation between our results and Ref.~\cite{Grzadkowski:2005yz}.
According to their results with the Scherk-Schwarz phase $\alpha$, in general, both the zero and nonzero KK modes become Majorana fermions.
This is because the global U(1) symmetry is lost due to the CBC; namely, none of the KK modes has a U(1) symmetry.
In fact, in their analysis, all of the mass eigenstates are described in terms of Majorana fermions whose masses are 
\begin{equation}
  m_n=\sqrt{m^2+\left(\frac{2\pi n+\alpha}{2\pi R}\right)^2}
\end{equation}
with an integer $n$. Here, $\alpha$ is the phase for the Scherk-Schwarz mechanism.
The two distinct Majorana masses $m_n$ and $m_{-n}$ become degenerate if the phase $\alpha$ is taken to be zero, 
and then our result is reproduced.
We also point out that there exists an accidental global U(1) symmetry in each KK mode if $\alpha=0$.
After the chiral rotation, 
the expression \eqref{eq:diagolalizedKK} shows that they are invariant under the SO(2) rotation in the basis of the two different Majorana fermions $\psi_\rmM'^{(n)}$ and $\psi_{\bar \rmM}'^{(n)}$. 
In other words, after composing the Dirac spinor $\Psi_\rmD^{(n)}=\frac{1}{\sqrt2}\left[\psi_\rmM'^{(n)}+i\psi_{\bar \rmM}'^{(n)}\right]$,
there is an accidental U(1) symmetry at each nonzero KK level.
On the other hand, in the zero-mode sector, the remaining Majorana fermion is 
$\psi_{\rmM}^{(0)}$ or $\psi_{\bar\rmM}^{(0)}$, so that there is no accidental U(1) symmetry.
This is why the nonzero KK modes become Dirac fermions.

Finally, we comment on the relation between the KK mass diagonalization and the axial U(1) transformation in this derivation.
As pointed out in the previous section, the CBC is compatible with the axial
U(1) transformation. However, this does not mean that the action itself
has an axial U(1) symmetry. In the present case, for example, the fermion
mass term explicitly breaks the axial U(1) symmetry.
The important point is instead that the axial U(1) transformation is
compatible with the CBC. The reason is that an eigenstate of the CBC
remains an eigenstate of the CBC even after the axial U(1) transformation.
Thus, since the KK mass eigenstates satisfy the Majorana condition as
a consequence of the CBC, the axial U(1) transformation does not mix
different KK modes. This also ensures that each KK mode remains a
Majorana fermion after the transformation.
Therefore, the axial U(1) transformation is suitable for field redefinition.
This is why such a transformation plays a crucial role in the diagonalization of the KK mass eigenstates.

\section{Extended seesaw scenario with the CBC and chiral orbifolding} 
\label{sec:SEESAW}

Now, we discuss the extended seesaw scenario as an application of the CBC.
As mentioned in the previous section,
the CBC can yield a Majorana fermion with a Majorana mass term in the zero mode.
Then, the simplest type-I seesaw 
is expected to be realized by utilizing the CBC.
Although this possibility is appealing, \emph{the naive extension is potentially problematic from a phenomenological point of view}, as we will discuss below.

The typical structure of the simplest type-I seesaw scenario in four-dimensional theory includes two interactions:
a Yukawa interaction and a Majorana mass term,
\begin{equation}
  y_\nu \bar L_L (i\sigma_2 H^\ast) N_R +\frac12 \mu\bar N_R N_R^\rmc.
\end{equation}
Here, $L_L$, $H$, and $N_R$ are the lepton doublet, Higgs doublet, and right-handed neutrino, respectively.
The parameters $y_\nu$ and $\mu$ stand for the four-dimensional Yukawa coupling and the large Majorana mass term for the neutrino in the sense of the seesaw scenario.
The Majorana mass term $\mu$ induces LNV, whereas the Yukawa interaction conserves lepton number,
as in the conventional type-I seesaw.
This is a nice feature; 
however,
the simplest type-I seesaw using the CBC induces tree-level LNV in the SM Yukawa coupling.
In fact,
this interaction originates from the following five-dimensional bulk interactions with the CBC and chiral orbifolding, as mentioned in the previous section:
\begin{equation}
  Y_\nu \bar L (i\sigma_2 H^\ast) (N+\gamma_5N^\rmc) +\mu\bar N N 
\end{equation}
where $Y_\nu$ stands for the five-dimensional Yukawa coupling, 
and we impose the following boundary condition:
\begin{equation}
\begin{bmatrix}
  L\\
  N\\
  H
\end{bmatrix}(y_i-y)
=
\begin{bmatrix}
  -\gamma_5 L\\
  N^{\rmc}\\
  H
\end{bmatrix}(y_i+y).
\end{equation}
As mentioned in the previous section, the first term clearly violates lepton number, so that the induced Yukawa coupling includes LNV.
This is potentially problematic from a phenomenological point of view.

Since the naive extension of the simplest type-I seesaw seems to have difficulties, the extended seesaw scenario, which includes two new different neutral fermions, is a candidate for a phenomenologically acceptable model.
In the conventional extended seesaw scenario in four-dimensional theory, there are two 
different singlet fermions, $N_{R}$ and $\Phi$,
and their interaction is written as 
\begin{equation}
\label{eq:interactionES}
  y_\nu \bar L_L (i\sigma_2 H^\ast) N_R +\mu' \bar N_R \Phi +m_{\Phi}\bar \Phi \Phi^\rmc .
\end{equation}
Since $L_{L}$, $N_{R}$, and $\Phi$ carry lepton number, the Majorana mass term \(m_{\Phi}\bar\Phi \Phi^{\rmc}\) clearly violates lepton number conservation
\footnote{The Majorana mass term in such a scenario is often assumed to be small by 't Hooft naturalness.}
.
These interactions can be embedded into the following bulk interaction in our scenario:
\begin{equation}
\label{eq:interactionES-CBC}
  Y_\nu \bar L (i\sigma_2 H^\ast) N 
  + M_N\bar N (\Phi-\gamma_5\Phi^\rmc) +M_\Phi\bar \Phi \Phi
\end{equation}
by imposing the CBC on $\Phi$. The other fermions obey the chiral orbifold projection,
and then the chiral fermions appear in the zero-mode sector.
After dimensional reduction, the zero mode realizes the extended seesaw structure \eqref{eq:interactionES}.
We emphasize that the Majorana mass term $M_\Phi$ in the third term
of \eqref{eq:interactionES-CBC}
conserves lepton number. This is because $\Phi$ does not carry lepton number.
However, the cross term $M_N$ in the  
second term induces LNV in contrast to the conventional extended seesaw scenario.
Thus, in our scenario, unlike in the conventional extended seesaw, LNV originates from the CBC-compatible mixing interactions rather than from a Majorana mass term.

\subsection{Our model}
In order to see how the extended seesaw scenario arises from the CBC, we provide a concrete model.
Since a Majorana fermion can be generated as a bulk fermion,
the extended seesaw scenario can be constructed without brane interactions
\footnote{
A similar interaction was pointed out in \cite{Saito:2010xj}; however, they introduce the SM particles as boundary fermions.
}.

We consider a Universal Extra Dimensions model (UED) in Ref.~\cite{Appelquist:2000nn} with two new singlet fermions, \(N\) and \(\Phi\).
The Lagrangian is given by 
\begin{equation}
\label{eq:Lagrangian}
  \begin{split}
    \mathcal L_\text{lepton}
    =&
    \bar L iD_M\Gamma^M L + \bar E iD_M\Gamma^M E +\bar N i\partial_M \Gamma ^M N
    +\bar \Phi (i\partial_M\Gamma^M -M_\Phi) \Phi
    \\
    &
    -Y_\nu \bar L (i\sigma_2H^\ast) N -Y_e \bar L H E- \frac{1}{\sqrt2} \bar N M_N(\Phi -\gamma_5\Phi^\rmc)+\text{(h.c.)}
  \end{split}
\end{equation}
where $L$, $E$, and $H$ stand for the SU(2) lepton doublet fermion $L=(n,e)^\rmT$, the singlet fermion for the right-handed electron, and the Higgs doublet, respectively.
$N$ and $\Phi$ are the right-handed neutrino and a new singlet neutral fermion, respectively.
$D_M$ is the corresponding covariant derivative of the SM gauge symmetry.
We impose the following boundary conditions on the fields:
\begin{equation}
  \begin{bmatrix}
    L\\E\\N\\H\\ \Phi
  \end{bmatrix}
  (y_i-y)
  =
  \begin{bmatrix}
    -\gamma_5L\\ +\gamma_5E\\ +\gamma_5N\\ H\\ \Phi^\rmc
  \end{bmatrix}
  (y_i+y).
\end{equation}
After integrating over the \(y\)-direction, \(L\), \(E\), and \(N\) give rise to chiral massless fermions, which behave as the chiral fermions of the SM. 
The field \(H\) behaves as the Higgs field, while \(\Phi\), subject to the CBC, appears as a four-dimensional Majorana fermion, as expected.
Table~\ref{tab:fieldscontents} summarizes the SM charges, boundary conditions, resulting zero modes, and lepton numbers assigned to these fields.
\begin{table}[h]
\centering
\caption{Field content and boundary conditions.}
\label{tab:fieldscontents}
\begin{tabular}{c|c|c|c|c}
\hline\hline
Field &
SU(2)$_L \times$U(1)$_Y$ &
type of the boundary condition &
Zero mode &
Lepton number 
\\
\hline
$L$
&
$(\mathbf 2,-1/2)$
&
chiral orbifolding
&
$L_L^{(0)}$
&
$1$
\\
$E$
&
$(\mathbf 1,-1)$
&
chiral orbifolding
&
$E_R^{(0)}$
&
$1$
\\
$N$
&
$(\mathbf 1,0)$
&
chiral orbifolding
&
$N_R^{(0)}$
&
$1$
\\
$H$
&
$(\mathbf 2,1/2)$
&
Neumann boundary condition
&
$H^{(0)}$
&
$0$
\\
$\Phi$
&
$(\mathbf 1,0)$
&
CBC
&
$\Phi_\rmM^{(0)}$
&
$0$
\\
\hline
\end{tabular}
\end{table}

We comment on our model in turn.
First, lepton number in this setup is not conserved by the non-trivial interaction between $\Phi$ and $N$ in \eqref{eq:Lagrangian}.
On the other hand, the induced Majorana mass term for $\Phi$ preserves lepton number because conventional lepton number cannot be assigned to $\Phi$.
Thus, these features are quite different from those of the conventional extended seesaw scenario.
In fact, from the lepton number assignments in Table~\ref{tab:fieldscontents},
the non-trivial interaction $\bar N M_N(\Phi -\gamma_5\Phi^\rmc)+\text{(h.c.)}$ breaks lepton number conservation.

Second, the mass parameters $M_N$ and $M_\Phi$ in our model depend on the UV scenario; however, their orders are naively expected to be much larger than the electroweak (EW) scale 
in the absence of any mechanism or symmetry.
This is because they are bulk interactions that preserve five-dimensional invariance and are consistent with the boundary conditions.
Thus, we assume that they are of the same order and much larger than the EW scale, \emph{i.e.}, $M_\Phi\sim M_N$, throughout this paper.

The mode expansions in this scenario are given by
\begin{equation}
 \begin{split}
  &H=\sum_{n=0}^\infty H^{(n)} C_n=\frac{1}{\sqrt{2\pi R}}H^{(0)}+\sum_{n=1}^\infty H^{(n)} C_n,\\
  &L=\sum_{n=0}^\infty\left[ L^{(n)}_L C_n+L^{(n)}_R S_n\right]=\frac{1}{\sqrt{2\pi R}}L_L^{(0)}+\sum_{n=1}^\infty\left[ L^{(n)}_L C_n+L^{(n)}_R S_n\right],\\
  &E=\sum_{n=0}^\infty\left[ E^{(n)}_R C_n+E^{(n)}_L S_n\right]=\frac{1}{\sqrt{2\pi R}}E_R^{(0)}+\sum_{n=1}^\infty\left[ E^{(n)}_R C_n+E^{(n)}_L S_n\right],\\
  &N=\sum_{n=0}^\infty\left[ N^{(n)}_R C_n+N^{(n)}_L S_n\right]=\frac{1}{\sqrt{2\pi R}}N_R^{(0)}+\sum_{n=1}^\infty\left[ N^{(n)}_R C_n+N^{(n)}_L S_n\right],\\
  &\Phi=\frac{1}{\sqrt2}\sum_{n=0}^\infty\left[ \Phi_{\rmM} ^{(n)} C_n+i\Phi_{\bar\rmM}^{(n)} S_n\right]
  =\frac{1}{2\sqrt{\pi R}}\Phi_{\rmM}^{(0)}+ \frac{1}{\sqrt2}\sum_{n=1}^\infty\left[ \Phi_{\rmM} ^{(n)} C_n+i\Phi_{\bar\rmM}^{(n)} S_n\right].
\end{split} 
\end{equation}
After electroweak symmetry breaking (EWSB), $\langle H\rangle =\frac{1}{\sqrt{2\pi R}}(0,v/\sqrt2)^\rmT$,
the quadratic terms become
\begin{equation}
  \begin{split}
    \mathcal L_\text{eff}
    =&
    \int_{-\pi R}^{\pi R}dy \mathcal L_\text{lepton}
    \\
    \supset&
    \bar L _L^{(0)} i\partial_\mu \gamma^\mu L_L^{(0)}
    +\bar E_R^{(0)} i\partial_\mu\gamma^\mu E_R^{(0)}
    +\bar N_R^{(0)} i\partial_\mu\gamma^\mu N_R^{(0)}
    +\frac12 \bar \Phi_\rmM^{(0)} (i\partial_\mu\gamma^\mu-M_\Phi) \Phi_\rmM^{(0)}
    \\
    &
    -\bar n_L^{(0)} y_\nu \frac{v}{\sqrt2} N_R^{(0)}-\bar e_L^{(0)} y_e \frac{v}{\sqrt2} E_R^{(0)} - \bar N_R^{(0)} M_N \Phi_\rmM^{(0)}
    +\text{(h.c.)}.
  \end{split}
\end{equation}
Extracting the neutrino sector in the zero mode, we have
\begin{equation}
  \begin{split}
    \mathcal L_\nu
    =&
    \bar n_L i\partial_\mu\gamma^\mu n_L
    +\bar N_R i\partial_\mu\gamma^\mu N_R
    +\frac12 \bar \Phi_\rmM (i\partial_\mu\gamma^\mu-M_\Phi) \Phi_\rmM
    \\
    &
    -\bar n_L M_\nu N_R- \bar N_R M_N \Phi_\rmM
    +\text{(h.c.)}
    \\
    =&
    \bar n_L i\partial_\mu\gamma^\mu n_L
    +\bar N_R i\partial_\mu\gamma^\mu N_R
    +\frac12 \bar \Phi_\rmM i\partial_\mu\gamma^\mu \Phi_\rmM
    -\frac12
    \overline{\Xi^\rmc} 
    M
    \Xi 
  \end{split}
\end{equation}
where $\Xi$ stands for $(n_L,N_R^\rmc, \Phi_\rmM)^\rmT$.
$M_\nu$ is the mass from the Yukawa interaction, $M_\nu = y_\nu v/\sqrt2$.
Hereafter, we concentrate on the zero mode, and KK indices such as $(0)$ on the fields are omitted.
We assume that $M_N$ and $M_\Phi$ are much larger than the EW scale,
which depends on the UV theory.
In fact, such an interaction is derived from the following interaction with a new SM-singlet charged scalar $S$:
\begin{equation}
 \frac{1}{\sqrt2} \kappa \bar N (S^\ast\Phi -S \gamma_5\Phi^\rmc)+\text{(h.c.)}
\end{equation}
where $\Phi$ and $S$ have the same charge, as mentioned in the previous section.
This interaction corresponds to Eq.~\eqref{eq:majoranainteraction} with
$\eta=N, \psi=\Phi, \phi=S$, and the even parity assignment for $S$.
After $S$ obtains a VEV, the above interaction yields the last term in the Lagrangian \eqref{eq:Lagrangian}.
In such a case, no lepton number can be assigned to the fields $S$ and $\Phi$, and the coupling constant $\kappa$ carries lepton number
\footnote{For another simple extension, an interaction $\kappa S\bar N (\Phi - \gamma_5\Phi^\rmc)+\text{(h.c.)}$ may be the UV completion, where $S$ is a scalar.
In such a case, lepton number can be assigned to both $S$ and $N$. After the scalar $S$ obtains a VEV, lepton number conservation breaks down spontaneously, and then the lepton-number-violating mass term $M_N$ is generated.}.

\subsection{Extended seesaw scenario and neutrino mass}
In this subsection,
we show how the neutrino mass is naturally explained.
The mass matrix in our model is very similar to that in the extended seesaw scenario.
Denoting the heavy sector by $M_\rmH$, we have
\begin{equation}
M=
     \begin{bmatrix}
      0&M_\nu^\rmT & 0\\
      M_\nu & 0&M_N\\
      0&M_N^\rmT&M_\Phi
    \end{bmatrix}
    =
    \begin{bmatrix}
    0&m_D^\rmT\\ m_D &M_\rmH
    \end{bmatrix}
\end{equation}
where $m_D^\rmT=(M_\nu^\rmT,0)$.
Since we assume the hierarchy $M_\nu\ll M_\Phi,M_N$,
the following rotation matrix can diagonalize the mass matrix up to order $\epsilon$:
\begin{equation}
\label{eq:rotationmatrix}
  R=\begin{bmatrix}
    1&-m_D^\rmT (M_\rmH)^{-1} \\ (m_D^\rmT (M_\rmH)^{-1})^\rmT &1
  \end{bmatrix}
  =\begin{bmatrix}
    1&\epsilon\\
    -\epsilon^\rmT&1
  \end{bmatrix}.
\end{equation}
Using the standard block-diagonalization, the block-diagonalized mass matrix becomes
\begin{equation}
  M_\text{Diag}
  =RMR^\rmT
  =\begin{bmatrix}
    \epsilon m_D & -\epsilon m_D\epsilon
    \\
    -\epsilon^\rmT m_D^\rmT\epsilon^\rmT & M_\rmH -m_D\epsilon -\epsilon^\rmT m_D^\rmT
  \end{bmatrix}.
\end{equation}
Then, the mass matrix for the active neutrino $m_\nu$ can be read off as
\begin{equation}
\begin{split}
  m_\nu
  =\epsilon m_D
  =M_\nu^\rmT (M_N^\rmT)^{-1}M_\Phi M_N^{-1}M_\nu.
\end{split}
\end{equation}
Now we assume that $M_\Phi,M_N\sim \mathcal O (10\text{TeV})$ and that $M_\nu$ is of the same order as the electron mass scale,
$\mathcal O(\text{MeV})$.
This assumption leads to a tiny neutrino mass,
\begin{equation}
  m_\nu =M_\nu^\rmT (M_N^\rmT)^{-1}M_\Phi M_N^{-1}M_\nu
  \sim \mathcal O (0.1{\rm eV}).
\end{equation}

Note that the above estimate relies on the smallness of the four-dimensional effective Yukawa coupling, $M_\nu\ll M_\Phi,M_N$.
In this paper, we assume that the four-dimensional effective Yukawa coupling in the neutrino sector is of the same order as those of the electron or first-generation quarks.
Ordinarily, such smallness of the Yukawa coupling is explained by the localization of mode-function profiles with respect to the extra dimension \cite{ArkaniHamed:1999dc}.
Such profiles are realized by introducing an odd-sign bulk mass for the chiral orbifolding fermions.
Thus, our statement is that such hierarchies in the Yukawa couplings in the charged-lepton or quark sectors are either assumed or explained by introducing an odd-sign bulk mass.
On the other hand, for the neutrino sector, 
the order of the neutrino Yukawa coupling is the same as that of the electron, which is explained by the localization of mode functions, 
while the extremely small mass is explained by the TeV-scale seesaw scenario, as discussed above.

We emphasize that the odd-sign bulk mass has limited effects on our scenario.
The localization of the profile gives a large suppression to the four-dimensional effective Yukawa couplings because the chiral fermions localize at opposite fixed points.
However, the non-trivial interaction between $N$ and $\Phi$ in the zero mode does not suffer from such suppressions.
This is because the zero mode of the CBC fermion $\Phi$ has a flat profile, as we derived in the previous section.
Therefore, for example, the typical zero-mode mixing between $N$ and $\Phi$ is given by
\begin{equation}
  M_N^{\rm eff}
  =
  M_N
  \int dy\, f_N(y) C_0(y)
  =
  M_N
  \int dy\, \sqrt{\frac{\mu_N}{\rme^{2\pi R\mu_N}-1}}\rme^{\mu_Ny} \frac{1}{\sqrt{2\pi R}},
\end{equation}
where $\mu_N$ stands for the odd-sign bulk mass for $N$.
Namely, the overlap integral between $N$ and $\Phi$ does not yield an exponential suppression factor.
Then, the assumption $M_N^\text{eff}\sim M_\Phi$ still holds after dimensional reduction.

Finally, we comment on the differences between our model and other scenarios. 
In general, extended seesaw scenarios introduce more than two neutral fermions in addition to the SM.
In our model, the mass matrix for the neutral fermions is described by 
\begin{equation}
M=
     \begin{bmatrix}
      0&M_\nu^\rmT & 0\\
      M_\nu & 0&M_N\\
      0&M_N^\rmT&M_\Phi
    \end{bmatrix}.
\end{equation}
In our model, we note again that the mixing term $M_N$ yields LNV, whereas the Majorana mass term $M_\Phi$ itself does not violate lepton number.
The inverse seesaw scenario is the most closely related framework to the present model.
In the conventional inverse seesaw scenario, the Majorana mass term generically provides the source of LNV.
Thus, the Majorana mass is assumed to be small, reflecting the smallness of LNV in the sense of the 't~Hooft naturalness criterion, and the tiny active neutrino mass is explained simultaneously~\cite{Mohapatra:1986aw}.
Although the neutrino mass matrix in our model is exactly the same as that of the inverse seesaw scenario, the origin of LNV is the mixing term $M_N$ in our model.
Therefore, the present framework does not require the Majorana mass $M_\Phi$ to be hierarchically smaller than the other heavy-sector mass parameters.
Accordingly, one may consistently consider the parameter region with $M_N\sim M_\Phi$ without imposing an additional hierarchy between these mass parameters.

\subsection{Phenomenological implications}
Finally, we discuss the viability of this model.
A precise flavor structure is necessary for comparison with experimental results;
we will discuss this in future work.
Instead, we give order-of-magnitude estimates for some observables.
First, the
unitarity violation is much smaller than the current upper bound.
The $\epsilon$ of the rotation matrix $R$ in \eqref{eq:rotationmatrix}, which connects the active neutrino and sterile neutrino, is given by 
\begin{equation}
  \epsilon=-m_D^\rmT (M_\rmH)^{-1} \sim \mathcal O(10^{-7}).
\end{equation}
The components corresponding to the active neutrino in the rotation matrix $R$ are roughly estimated as
$1-\epsilon^2$.
Then, the unitarity violation becomes
$\epsilon^2\sim \mathcal O(10^{-14})$,
which is much smaller than the current observational bound, $10^{-3}-10^{-4}$ \cite{Fernandez-Martinez:2016lgt}.
In our model, the smallness of the unitarity violation originates from the assumption that the four-dimensional effective neutrino Yukawa coupling is comparable to that of the first-generation charged lepton, corresponding to
\(M_{\nu} \sim \mathcal{O}(1\,\mathrm{MeV})\).
As discussed in the previous section, this can be realized by introducing odd-sign bulk masses for the chiral orbifolding fermions \(L\), \(E\), and \(N\).
If, instead, one assumes that the four-dimensional effective neutrino Yukawa coupling is comparable to that of the third-generation charged lepton, corresponding to
\(M_{\nu} \sim \mathcal{O}(1\,\mathrm{GeV})\),
a somewhat larger hierarchy between \(M_N\) and \(M_{\Phi}\) is required to realize the tiny active neutrino mass.
Since an odd-sign bulk mass term mixes different KK modes, such mixing may give large contributions to other observables.

We briefly comment on the electric dipole moment (EDM) of the electron.
In Ref.~\cite{EuropwanEDMprojects:2025okn}, the mass range testable by EDM measurements is $10$--$100\,\rm TeV$.
This suggests that TeV-scale seesaw scenarios, including our model, are testable.
However, due to the smallness of the mixing between active and sterile neutrinos, a large suppression may appear in the electron EDM.
In fact, the EDM of the electron in this setup is roughly estimated as
\begin{equation}
  d_e\sim e\frac{m_e}{(16\pi^2)^2M_\rmH^2}\epsilon^2\sim \mathcal O (10^{-44})e\cdot\rm{cm}.
\end{equation}
It is also much smaller than the current observational bound $d_e<0.041\times 10^{-28}e\cdot\rm{cm}$ summarized by the Particle Data Group~\cite{ParticleDataGroup:2024}.

The above order estimates assume that the small neutrino Yukawa coupling, $\mathcal O(10^{-5})$, is simply taken as an input parameter.
In this setup, there are no KK-mode mixings induced by terms such as an odd-sign bulk mass.
Then, the contributions from the nonzero KK modes are expected to be subdominant. 

On the other hand, if the small effective Yukawa coupling is realized by the localization of the profile caused by the odd-sign bulk mass term,
the situation may be different.
In such a case, the Yukawa coupling suffers from exponential suppression due to the localization of the profile.
However, the Yukawa coupling in the nonzero KK-mode sector may be relatively large.
Therefore, the nonzero KK-mode contributions are not negligible.

\section{Summary and discussion} 
In this paper, we discuss the structure of the KK mass eigenstates under the CBC, which is known for realizing a Majorana fermion within a five-dimensional theory.
By solving the equation of motion of the free theory, we show that the nonzero KK modes $\psi_\rmM^{(n)}$ and $\psi_{\bar\rmM}^{(n)}$ have the same mass $m_n=\sqrt{m^2+\left(\frac{n}{R}\right)^2}$.
We obtain this result by direct calculation and by utilizing the axial U(1) transformation.
The degeneracy of the two independent Majorana fermions in each nonzero KK mode implies that they are described by a single Dirac spinor, although the zero mode is still a Majorana fermion.
In fact, as we show in the main text, there is an accidental U(1) symmetry at each KK level, although the global U(1) symmetry of the five-dimensional fermion $\psi(y)$ is lost due to the CBC.

We note that this accidental U(1) symmetry does not exist in the zero-mode sector because either $\psi_\rmM^{(0)}$ or $\psi_{\bar\rmM}^{(0)}$ is projected out.
We also note that the above degeneracy may be lifted by interactions, such as gauge interactions.
In fact, as pointed out in Ref.~\cite{Grzadkowski:2005yz}, the nonzero KK mass eigenstates of the fermion in the U(1) gauge theory 
split into two different mass eigenvalues by the VEV of $A_y$, or equivalently by the Scherk-Schwarz mechanism.

The compatibility between the axial U(1) transformation and the CBC plays a crucial role in the diagonalization of the KK mass eigenstates,
\emph{i.e.}, each KK mode $\psi_\rmM^{(n)}$ and $\psi_{\bar\rmM}^{(n)}$ is still a Majorana fermion after the axial U(1) transformation.
We also emphasize that the action may not preserve the axial U(1) symmetry; however, this compatibility guarantees that the axial U(1) transformation is valid 
in the diagonalization of the KK mass eigenstates.
This is because the KK modes are eigenstates of the boundary condition (CBC), and they are still eigenstates of the CBC after the axial U(1) transformation.

Allowed interactions between fermions with the CBC and chiral orbifold projection are also discussed.
The quadratic interactions $\bar \eta \psi$ and $\bar \eta\gamma_5\psi^\rmc$ are not compatible with the boundary conditions;
however, a linear combination of them, such as $\bar \eta (\psi\pm \gamma_5\psi^\rmc)$, agrees with the CBC and chiral orbifolding.
We note that $\eta$ and $\psi$ are neutral because a definite fermionic charge cannot be assigned to $\psi$ by the CBC. 
We also propose Yukawa-type interactions.
There are two kinds of Yukawa-type interactions:
one is an interaction with a neutral scalar $S$, such as $S\bar \eta (\psi\pm \gamma_5\psi^\rmc)$.
The other is an interesting case, an interaction with a charged scalar $S$, such as $\bar \eta (S^\ast\psi\pm S\gamma_5\psi^\rmc)$.
The charges of both $S$ and $\psi$ are the same, and $S$ obeys a suitable CBC.
In such a case, gauge interactions can be introduced for the charged scalar $S$ and $\psi$, as we will discuss in a forthcoming paper.

As an application of the above quadratic interaction, we discuss the possibility of the seesaw scenario by utilizing the CBC. 
The simplest type-I seesaw scenario can be constructed by the above interaction; however, 
the induced Yukawa coupling contains LNV, so that it is potentially dangerous from a phenomenological point of view.
Then, we focus on the extended seesaw scenario.
We provide a concrete model and show that this scenario can be phenomenologically viable, as we point out in the main text.
The LNV in our model is different from that in the conventional extended seesaw scenario;
namely,
the interactions between the neutral singlet fermions induce LNV rather than the Majorana mass term.
It would be interesting to further investigate its phenomenological implications in future work.
Finally, we discuss unitarity violation.
Under our assumption, $M_N\sim M_\Phi\sim \mathcal O(10\text{TeV})$ gives a relatively small unitarity violation, 
$\epsilon \epsilon^\dag\sim 10^{-14}$.
The smallness of the non-unitarity is not a generic prediction of extended seesaw models. 
It follows from our assumption
\(M_\Phi\sim M_N\) and \(M_\nu\sim{\cal O}({\rm MeV})\), for which
\(\epsilon\sim M_\nu/M_N\sim10^{-7}\).
Such smallness originates from the smallness of the neutrino Yukawa coupling, $y_\nu\sim \mathcal O(10^{-5})$.
In this setup, KK-mode mixing does not exist, so that the contributions from nonzero KK modes can be ignored.
We also pointed out that nonzero KK modes may give non-negligible contributions when the small neutrino Yukawa coupling is explained by
fermion localization.

\subsection*{Acknowledgments}

Y. Abe greatly appreciates valuable discussions with Y. Kawamura.
Y. Abe is supported by JSPS Grants-in-Aid for Early-Career Scientists (Grant No. JP23K13108).
Y. Fujimoto is supported by JSPS KAKENHI Grant Number JP23K03416.

\providecommand{\href}[2]{#2}\begingroup\raggedright\endgroup

\end{document}